# Real-Time Performance Analysis of Infrastructure-based IEEE 802.11 Distributed Coordination Function

Feng Xia*, Ruixia Gao, Linqiang Wang, Ruonan Hao

*School of Software, Dalian University of Technology, Dalian 116620, China*
*(e-mail: f.xia@ieee.org)*

**Abstract:** With the increasing popularity of wireless networks, wireless local area networks (WLANs) have attracted significant research interest, which play a critical role in providing anywhere and anytime connectivity. For WLANs the IEEE 802.11 standard is the most mature technology and has been widely adopted for wireless networks. This paper analyzes real-time performance of the IEEE 802.11 standard that adopts the MAC protocol of Distributed Coordination Function (DCF) operating in infrastructure mode. Extensive simulations have been done to examine how the network performance in terms of real-time metrics including effective data rate, latency and packet loss rate will be impacted by some critical parameters (e.g. CWmin and packet payload). The results are presented and analyzed. The analysis of simulation results can provide support for parameter configuration and optimization of WLANs for real-time applications.

## 1. INTRODUCTION

As no cables are required, wireless local area networks (WLANs) have become an essential amenity applied to more and more fields such as healthcare, manufacturing, academic areas, consumer appliances and so on. They can be deployed in hot spots areas and offer performance comparable to wired local area networks. The question of the performance effectively perceived by mobile hosts becomes increasingly important as many new emerging applications such as video surveillance, video conferencing and voice services, real-time multimedia communications require sufficient bandwidth to meet the need to address larger coverage areas, high capacity and low latency services, and an ever increasing number of users. In WLANs, the family of IEEE 802.11 protocols is perhaps the most widely adopted standard (IEEE Std 802.11b, 1999; IEEE Std 802.11g, 2003; IEEE Std 802.11e, 2005; Kim, 2005). Therefore, many challenges of the wireless medium are addressed by research especially to improve the real-time performance of the 802.11 DCF (Distributed Coordination Function), which is the basic operation of the medium access control (MAC) defined in 802.11.

The IEEE 802.11 standard provides detailed medium access control (MAC) and physical layer (PHY) specification for WLANs (IEEE 802.11 Working Group, 1997). The MAC incorporates two medium access methods, Distributed Coordination Function (DCF) and Point Coordination Function (PCF) (IEEE 802.11 Working Group, 1997; Crow et al., 1997; Hara and Petrick, 1999; Kumar et al., 2006). DCF is an asynchronous data transmission function, which is best suited to delay insensitive data and supports the infrastructure and ad-hoc configuration. On the other hand, the optional Point Coordination Function (PCF) is a synchronous data transmission function, which is a centralized MAC protocol able to support collision free and time bounded services, but it only supports the infrastructure configuration.

DCF, as the fundamental mechanism to access the medium, is a random access scheme, based on the carrier sense multiple access with collision avoidance (CSMA/CA) protocol. It describes two techniques to employ for packet transmission. The default scheme is a two-way handshaking technique called basic access mechanism. In addition to the basic access, an optional four way handshaking technique, known as request-to-send/clear-to-send (RTS/CTS) mechanism, is suited to combat the so-called problem of Hidden Terminals and reserve the medium when large packets are transmitted in order to reduce the duration of a collision (Huang and Chen, 1995).

This paper focuses only on DCF in the infrastructure mode which includes an access point (AP) and a certain number of stations. The simulation transmission scenarios apply an IEEE 802.11b infrastructure-based WLAN in which each station transmits to the AP a traffic flow using the basic CSMA/CA scheme. The used system parameters are adopted from the IEEE 802.11b standards (IEEE 802.11a/b, 1999). Effective data rate, delay, and packet loss rate are three performance metrics that are of great interest to real-time applications. Using the proposed model, an extensive performance evaluation of DCF is carried out by varying the system parameters such as the minimum contention window CWmin, packet payload, packet generation interval and bitrate. This paper contributes to the better understanding of the IEEE802.11 protocol by presenting a set of results of simulation experiments based on OMNet++ which is a popular public-source simulation platform especially suitable for the simulation of communication networks. These results allow us to determine how the protocol performance is affected by the system parameters.

---
*Corresponding author: Feng Xia

The remainder of this article is organized as follows. Section 2 gives an overview of related work in the literature. Subsequently, Section 3 presents briefly the IEEE 802.11 DCF. This is followed by simulation settings including simulation scenario and parameter settings, and definition of performance metrics in Section 4. Section 5 analyzes the simulation results. Finally, Section 6 concludes the paper.

## 2. RELATED WORK

In previous research, a number of researches have been done to study the performance on IEEE 802.11 especially on DCF mechanism. The throughput, frame discard probability, and average frame delay performance of DCF in IEEE 802.11 are analyzed and evaluated for different parameters in (Chen and Liu, 2010). In (Khayyat et al., 2007) the author presents a Markov chain analysis for studying the performance of the IEEE 802.11 Distributed Co- ordination Function. In (Amjad and Shami, 2006), the authors propose a new MAC protocol that modifies the DCF protocol such that the channel utilization can be improved with successful packet transmissions, yielding higher throughput performance.

Certain research papers aim to propose new analytical models for investigating the impact of some factors on the performance of IEEE 802.11. In (Kumar and Krishnan, 2010), the authors present an analytical model to analyze the saturation performance of IEEE 802.11 MAC layer with capture effects. Zhang and Zhao (2010) present a new model adding idle backoff process, under non-saturated traffic assumption. In (Zheng et al., 2006), the authors provide a new analytical model that considers the channel error conditions in the unsaturated mode to analysis delay and throughput performance of IEEE 802.11 DCF under different incoming traffic conditions. He et al. (2009) give an analytical model to evaluate the impact of background traffic on the performance of 802.11 power saving mechanism for unicast services.

In addition, some researchers study the protocol with focus on certain problem or specific applications which have particular requirements. Like in (Hung and Marsic, 2007), the authors evaluate the hidden station effect on the access delay of the IEEE 802.11 DCF in both non-saturation and saturation condition. In (Tian and Tian, 2010), the authors focus on the IEEE 802.11 with DCF for soft-real-time control application and quantitatively evaluate the network quality-of-service (QoS) performance in periodic real-time traffic environments. In (Bianchi, 2000), Bianchi concentrates on evaluating the throughput performance of IEEE 802.11 DCF, under the assumption of ideal channel conditions and finite number of terminals. Ivanov et al. (2010) consider the 802.11 networks which are defined in terms of throughput requirements and packet loss probability limitations and the influence of sizes of packets being transmitted through the network on the QoS is investigated. Xiao and Yin (2010) model and analyze the performance of IEEE 802.11 MAC protocol in multi-hop ad hoc networks. Jeong et al. (2010) propose a novel analysis based on simple mathematical model for the hidden node problem. Yang et al. (2009) analyze the goodput of a WLAN with hidden nodes under a non-saturated condition.

This paper has two key contributions. First, it extensively studies the real-time performance of the infrastructure mode of IEEE 802.11 DCF basic access mechanism based IEEE 802.11b, using the OMNeT++ simulator. It selects RTT (Round Trip Time), effective data rate, and packet loss rate as the network performance metrics and analyzes how they will be affected by several important protocol parameters. Second, this paper makes an in-depth analysis of the results, which can provide support to configure and optimize the parameters of IEEE 802.11 WLAN for real-time applications.

## 3. IEEE 802.11 DCF

As mentioned previously, IEEE 802.11 DCF includes two mechanisms for packet transmission, i.e. basic access mechanism and request-to-send/clear-to-send (RTS/CTS) access mechanism. This section briefly summarizes the two mechanisms. For a more complete and detailed presentation, please refer to the 802.11 standard (IEEE 802.11 Working Group, 1997).

### 3.1 The IEEE 802.11 Basic Access Method

The basic access mechanism is illustrated in Fig. 1. When the sender wishes to transmit a new packet, it needs to monitor the channel first. If the channel is sensed to be idle for a period of time equal to the Distribute Inter Frame Space (DIFS), it begins the transmission. Otherwise, if the channel is busy (either immediately or during the DIFS), the station persists to monitor the channel until it is measured idle for a DIFS. At this point, the station will generate a random number of backoff slot times between 0 and the backoff window size (CW) before transmitting, to minimize the probability of collision with packets being transmitted by other stations. In addition, to avoid channel capture, a station must wait a random backoff time between two consecutive new packet transmissions, even if the medium is sensed idle in the DIFS time. During the backoff procedure if the channel is idle for a DIFS interval, then the backoff timer is decremented as long as the channel is sensed idle, "frozen" when a transmission is detected on the channel and resumed when the channel is sensed idle again for more than a DIFS. When the backoff timer reaches to zero the station starts the transmission.

When the receiver receives the packet correctly, it waits for a short inter-frame space(SIFS) interval and then transmits a MAC acknowledgement(ACK) back to the sender to inform the successfully reception of the data packet. In case the sender does not receive the ACK, it considers the previous transmission is failed and schedules to retransmit the data packet. If the data packet is not transmitted successfully, the backoff window CW would be doubled until it reaches the maximum value CWmax. The backoff window would be reset to CWmin whenever a data packet is transmitted successfully, signified by an acknowledgement packet (ACK) or after retry limit unsuccessful attempts when the data packet is discarded.

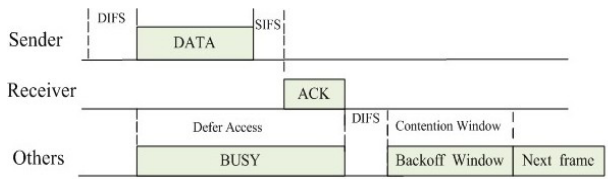

Fig. 1. The IEEE 802.11 basic access method

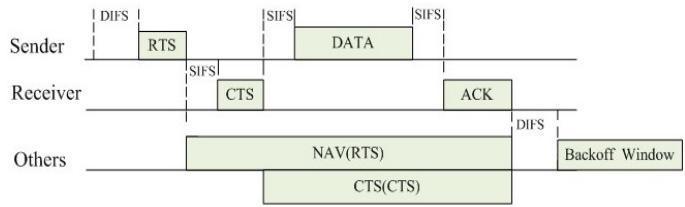

Fig. 2. The IEEE 802.11 RTS/CTS access method

*3.2 The IEEE 802.11 RTS/CTS Access Method*

The RTS/CTS access method is an additional four-way handshaking technique and very effective in solving the hidden terminal problem. The RTS/CTS mechanism is shown in Fig. 2. When the sender wants to transmit a packet, it sends a short frame called request to send (RTS) instead of the packet first after the channel has been sensed idle for a DIFS. When the receiver detects the RTS, it responses, after a SIFS, with a clear to send (CTS) frame. A successful RTS/CTS exchange reserves the channel for the sender-receiver pair. Other stations adjust their Network Allocation Vectors (NAVs) based on the duration field of the RTS or of the CTS. The sender starts to transmit the packet after a SIFS only if it received the CTS frame correctly. Same as basic access method, the receiver will send back an ACK to acknowledge after received the packet successfully. If the CTS is not received within a given time frame, the sender retransmits the RTS according to the backoff rules similar to the basic access method. If the RTS is not transmitted successfully after retry limit attempts, it is discarded altogether with the data packet following the backoff window being reset to CWmin.

The CTS and RTS frames contain information about the source, destination and duration required by the following transaction. This information can be read by any listening station. The stations overhearing either RTS or CTS frame adjust their NAVs to the duration specified in RTS/CTS frame. In this way, they can suitably delay further transmission and thereby avoid collision. Therefore, the RTS/CTS mechanism can combat the hidden terminal problem and increase the system performance by reserving the medium to reduce the duration of a collision when long messages are transmitted.

## 4. SIMULATION MODEL SETTINGS

This section describes the configuration and settings of the simulation model in OMNeT++, including simulation scenario and parameter settings, as well as definition of performance metrics. The model implements an IEEE 802.11b infrastructure WLAN using the basic CSMA/CA scheme and assumes that the network consists of $n$ contending stations transmitting in ideal conditions (no errors occur in the channel and no hidden stations exist). So the following assumptions are made: (1) Packets are lost only due to collision; (2) The hidden terminal effect is ignored.

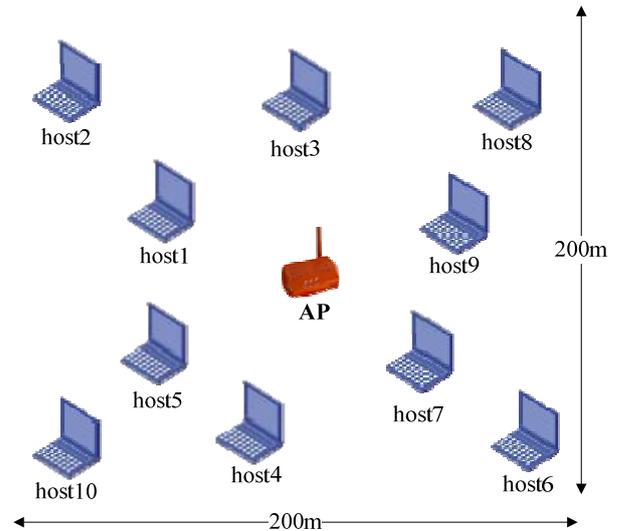

Fig. 3. Network topology

As mentioned in the previous section, this paper focuses on the infrastructure mode which includes an AP and a certain number of stations, as shown in Fig. 3. It consists of an AP and $n$ hosts. These hosts are randomly distributed in the 200m*200m playground with the AP at the centre of the playground. The transmission range of every node is 300m. Accordingly, all nodes are set to be in each other's radio range. These hosts will move in a speed which is the absolute value of a normally distributed random number, with an average of 20mps, and standard deviation of 8 mps. They periodically generate a packet randomly addressed to one of other hosts and forwarded by the AP.

Several critical parameters are selected as impact factors to be examined, including packet payload, packet generation interval, bit rate, and CWmin. These parameters will be introduced with scenarios in the next section. Some fixed important model parameters and default values of these parameters are listed in Table 1.

As mentioned in Section 1, to support more demanding applications the performance of WLANs needs to be real-time, reliable, and resource efficient. In order to meet these requirements, RTT, effective data rate and packet loss rate are selected as performance metrics since they are among the most popular metrics for evaluating network QoS.

Table 1. Parameter settings

| Playground size | 200*200 m*m |
|---|---|
| Transmitter power | 9.0 mW |
| Carrier frequency | 2.4 GHz |
| Carrier sense sensitivity | -85 dBm |
| PHY header | 192 bits |
| MAC header | 224 bits |
| Slot time | 20 μs |
| SIFS | 10 μs |
| DIFS | 50 μs |
| Transmission range | 300 m |
| macMTU | 1500 B |
| Number of packets sent by every host | 500 |
| Retry limit | 7 |
| CWmax | 1023 |
| Movement speed | \| normal(20 mps,8 mps) \| |
| Packet generation interval | 0.1 s (default) |
| CWmin | 31 (default) |
| Bit rate | 2 Mbps (default) |
| Packet payload (P) | 512 B (default) |

- *RTT:* it is a crucial metric to evaluate the real-time performance of networks. It refers to the average time difference between the points when a packet is sent and when an acknowledgment of that packet to be received, and the RTT time is also known as the ping time.

- *Effective data rate:* it is an important metric to evaluate the link bandwidth utilization which reflects the resource efficiency as well as dependability of networks. It is defined as below:

  $R_{effData} = N_{susspacket} \times P / (T_{end} - T_{start})$.

  where $N_{susspacket}$ is the total number of usable data packets which are transmitted successfully in the simulation time. $P$ is the size of the data packet payload. $T_{end} - T_{start}$ is the total time of the transmission from beginning to end.

- *Packet loss rate:* it indicates the performance of reliability, thus being an important metric. It is the ratio of the number of packets dropped by the network to the total number of packets sent by all hosts.

From the above definitions, it is easy to find that the effective data rate is closely related with the packet loss rate. Higher packet loss rate leads to lower effective data rate for the same number of transmitters.

## 5. SIMULATION RESULTS AND ANALYSIS

The previous section has described the common settings for the simulations. This section will analyze the impact of five impact factors (i.e. packet payload, packet generation interval, bit rate, and CWmin) on the performance of IEEE 802.11 WLANs in terms of the above mentioned metrics, respectively. During the process of simulation, when a specific parameter is examined as the impact factor, other parameters take the default values.

### 5.1 Impact of Packet Payload

Packet payload is the payload size of the data packet. It varies from 64 to1024 bytes, in terms of 64, 128, 256, 512, and 1024. Fig. 4 shows its influence on the performance metrics as a function of the number of hosts.

Figs. 4 (a), 4(b) and 4(c) present the effective data rate, packet loss rate and RTT with varied values of packet payload, respectively. As can be seen from Fig. 4 (a), the effective data rate increases as the number of hosts for 64 bytes payload. As for larger packet payloads like 128, 256 and 512 bytes, the performance degrades when more hosts transmitting, especially for the payload of 1024 bytes. From this analysis, it is easy to conclude that packet collision is more severe when the packet size is large. On the other hand, as shown in Fig. 4(b), more hosts and larger packet payload lead to higher packet loss rates. This is because as the number of hosts increases, a larger packet payload will has more possibility to collide and to be destroyed. Fig. 4(c) indicates that large packet payload like 1024 bytes incurs more processing delays as compared to small packet payload, e.g. 64 and 128 bytes. This can be explained that more hosts and larger packet payload increase the probability of packet collision. This can increase times of backoff and retransmission, which are a considerable factor for longer delay. Therefore, as shown in Fig. 4(c) the RTT grows as the increase of the number of hosts and packet payload.

### 5.2 Impact of Packet Generation Interval

All hosts periodically generate a packet randomly addressed to one of other hosts. The time interval between two adjacent packets' generation is referred to as packet generation interval. It is apparent the packet generation interval is inversely proportional to traffic load. The results are shown in Fig. 5.

Figs. 5(a), 5(b) and 5(c) show the effective data rate, packet loss rate and RTT with different packet generation intervals respectively. The results show that for small packet generation intervals which are less than 0.25 s, with an increasing number of hosts, the effective data rate first grows and then decreases, while the packet loss rate and the RTT exhibit a significant increase. As for larger packet generation interval like 0.5 and 1s, the effective data rate always keeps increasing while the packet loss rate and the RTT are close to zero and change hardly as the number of hosts increases. These effects can be explained as follows: the packet generation interval is inversely proportional to traffic load. Smaller packet generation interval means heavier traffic load which increases the probability of packet collision. This is the reason for the phenomenon when the packet generation

interval is less than 0.25s. In contrast, larger packet generation intervals imply lighter traffic load and hence few collisions happen. This is why the performance is fine for larger packet generation interval like 0.5 and 1s.

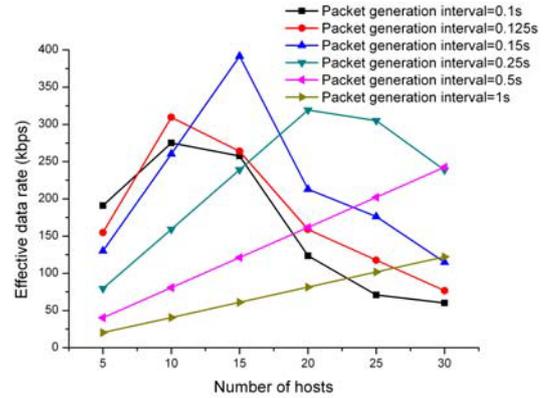

(a) Effective data rate

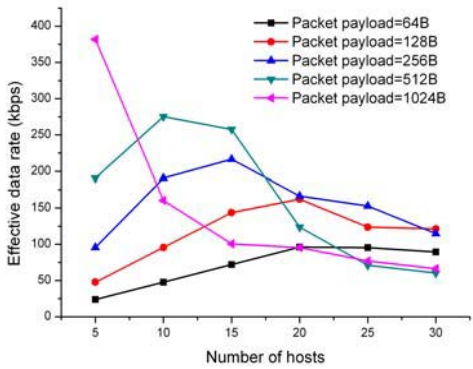

(a) Effective data rate

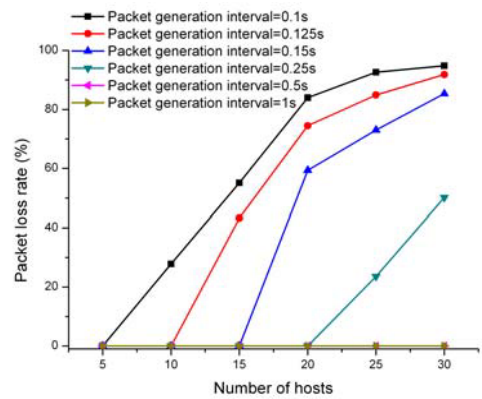

(b) Packet loss rate

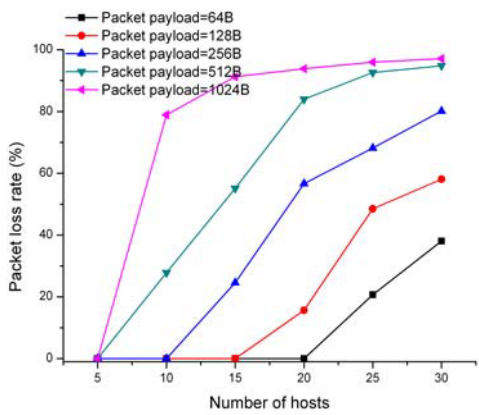

(b) Packet loss rate

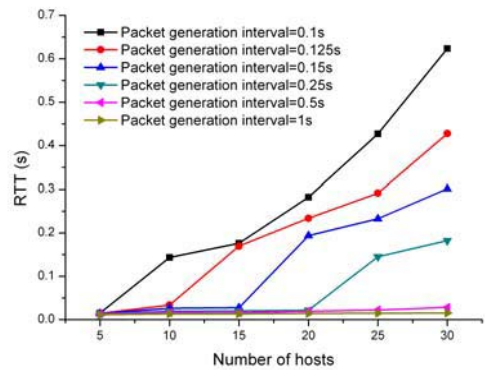

(c) RTT

Fig. 5. Performance with different packet generation intervals

### 5.3 Impact of Bitrate

Bitrate (a.k.a. bit rate) is the number of bits that are conveyed or processed per unit of time, and is also known as data rate or bandwidth. Fig. 6 shows the performance metrics as a function of the number of hosts for different data rates of 802.11b: 11 Mbps, 5.5 Mbps, 2 Mbps and 1 Mbps.

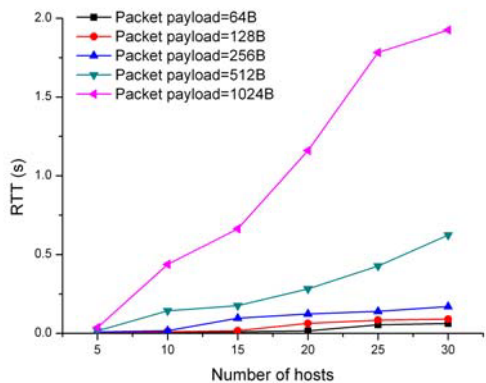

(c) RTT

Fig. 4. Performance with different packet payloads

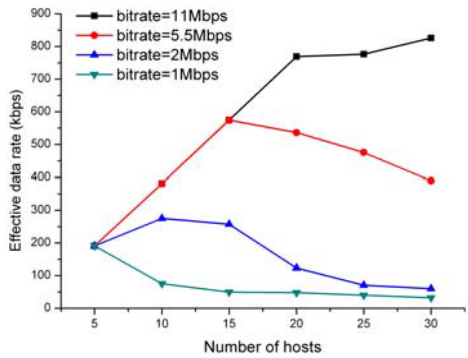

(a) Effective data rate

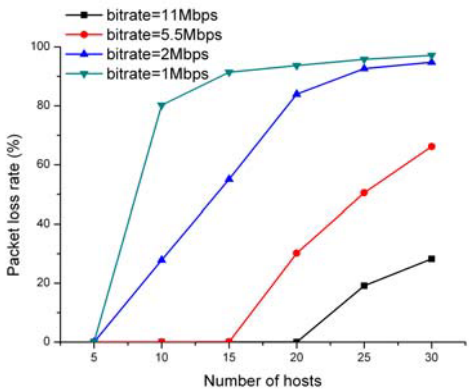

(b) Packet loss rate

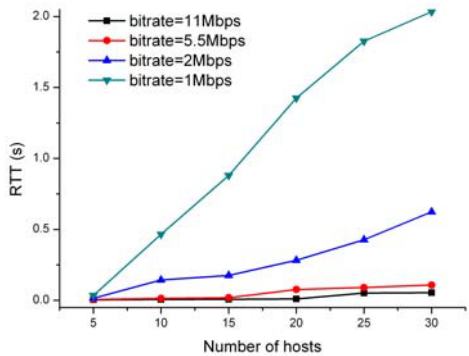

(c) RTT

Fig. 6. Performance with different bit rates

Figs. 6(a), 6(b) and 6(c) depict the measured effective data rate, packet loss rate and RTT with different bit rates, respectively. As shown in Fig. 6(a), when fewer (e.g. 5) hosts compete to access the channel, the effective data rate is the same for various bit rates. However, as the number of hosts increases, the effective data rate increases for large bitrate of 11 Mbps: it first grows and then decreases for 5.5 and 2 Mbps, and slumps for a small bitrate of 1 Mbps. This is because small bit rates only can satisfy bandwidth requirements for a few hosts. As the number of hosts increases, the bandwidth reaches saturation and the possibility of packet collisions increases. This is also the reason for the results of the packet loss rate and the RTT as shown in Figs. 6(b) and 6(c). It is clear that, for a small bit rate of 1 Mbps, as the number of contending hosts increases, the packet loss rate and the RTT increase precipitously. However, for larger bit rates like 11 and 5.5 Mbps, the packet loss rate and RTT perform well when more hosts transmitting.

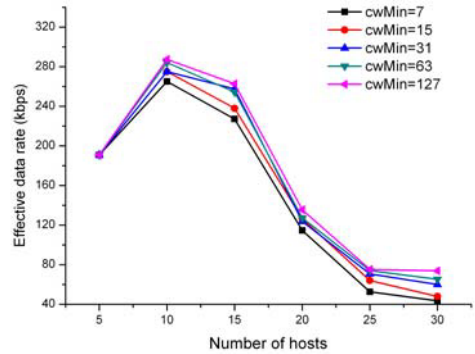

(a) Effective data rate

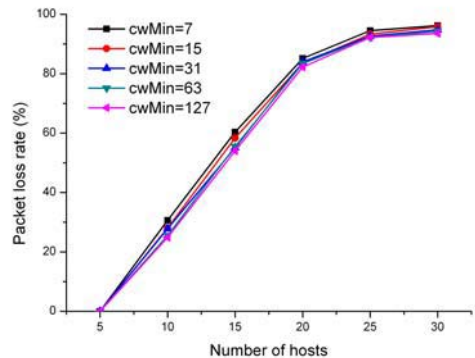

(b) Packet loss rate

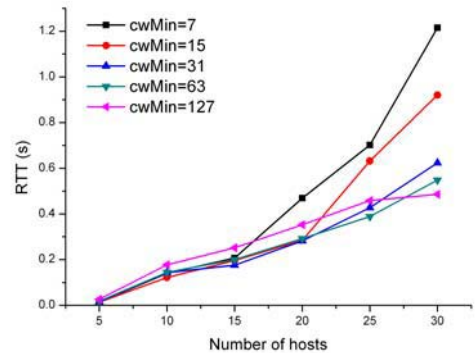

(c) RTT

Fig. 7. Performance with different CWmin values

## 5.4 Impact of CWmin

CWmin is the initial value of backoff window. Its default value is 31 in IEEE 802.11b. Fig. 7 shows the performance as a function of the number of hosts for different CWmin values.

As shown in Fig. 7(a), as the number of hosts increases, the trends of curves of the effective data rate are almost the same for varied CWmin, and so are those of the curves of the packet loss rate shown in Fig. 7(b). For fewer hosts like the number of 5, the effective data rate and packet loss rate almost stay the same for different CWmin values, while for more hosts the effective data rate grows slowly and the packet loss rate slumps slightly as CWmin increases. This is explained by the fact that there are more slot times available as CWmin gets larger, so it is less likely for two stations to have their counter equal to zero at the same time, which reduces the packet collisions. Fig. 7(c) reports the measured RTT. When fewer hosts like 5 or 10 compete to access the channel, the RTT grows slightly as CWmin increases. However, as the number of hosts increases, the RTT for small CWmin (like 7) increases precipitously, while that for large CWmin increases slowly. The reason for this may be that larger CWmin means more backoff slot times can be selected. Therefore, for fewer hosts this can increase longer backoff time which may incur larger RTT, on the other hand, for more hosts which can increase the possibility of detecting an idle channel.

To summarize the performance analysis in this section, the results show that the effective data rate, packet loss rate, and RTT are sensitive to the system parameters being chosen, proving the importance of dynamic tuning of the system parameters in IEEE 802.11. Moreover, the optimal values of the parameters are inconsistent with each other for different performance metrics, implying that some tradeoff among the metrics must be made in the dynamic tuning.

## 6. CONCLUSIONS

This paper carried out an insightful performance evaluation of IEEE 802.11 standard in the infrastructure mode. Considering general requirements of more demanding applications, several network performance metrics including effective data rate, packet loss rate, and RTT have been examined with respect to some important and variable protocol parameters. The analysis of simulation results demonstrate the importance of performing dynamic tuning of the system parameters in IEEE 802.11 and provides some insights for configuring and optimizing the IEEE 802.11 protocol for real-time applications.

Our future work will examine how to extend/modify IEEE 802.11 to make it more suitable for high QoS applications. Self-adaptive and autonomous approaches will be our focus.

## ACKNOWLEDGMENTS

This work was partially supported by the Natural Science Foundation of China under Grant No. 60903153, the Fundamental Research Funds for Central Universities (DUT10ZD110), the SRF for ROCS, SEM, and DUT Graduate School (JP201006).